# Strong Phonon Anharmonicity of Clathrate Compound at High Temperature

Masato Ohnishi[1,*], Terumasa Tadano[2,3], Shinji Tsuneyuki[4,5], and Junichiro Shiomi[1,*]

[1]*Department of Mechanical Engineering, The University of Tokyo, 7-3-1, Hongo, Bunkyo-ku, Tokyo 113-8656, Japan*

[2]*Research Center for Magnetic and Spintronic Materials, National Institute for Materials Science (NIMS), 1-2-1 Sengen, Tsukuba, Ibaraki 305-0047, Japan*

[3]*Elements Strategy Initiative Center for Magnetic Materials, National Institute for Materials Science (NIMS), 1-2-1 Sengen, Tsukuba, Ibaraki 305-0047, Japan*

[4]*Department of Physics, The University of Tokyo, 7-3-1, Hongo, Bunkyo-ku, Tokyo 113-0033, Japan*

[5]*Institute for Solid State Physics, The University of Tokyo, 5-1-5, Kashiwanoha, Kashiwa, Chiba 277-8581, Japan*

Effects of strong phonon anharmonicity of a type-I clathrate $Ba_8Ga_{16}Sn_{30}$ induced by quadruple-well potential of guest atoms were investigated. Phonon transport including coherent interbranch component was analyzed using a first-principles-based self-consistent phonon (SCP) theory that gives temperature-dependent harmonic interatomic force constants and by solving off-diagonal components of group velocity operator. Experimentally observed thermal conductivities have been reasonably reproduced by considering both lattice and electron contributions. Through the analysis with the SCP theory, we found that hardening of guest modes leads to an increase in lattice thermal conductivity at frequencies below those of framework-dominant flat modes (< 40 cm$^{-1}$), which finally results in the slow decay and slight increase in the total lattice thermal conductivity with increasing temperature. Detailed analyses revealed that the increase in lattice thermal conductivity at low frequency is attributed to (a) the increase in group velocities of phonon modes located at frequencies below that of the flat guest modes and (b) abnormal increase in lifetimes of phonon modes located between frequencies of the flat guest and framework modes with increasing temperature. From an engineering point of view, this effect may lead to an intriguing phenomenon, a larger decrease in thermal conductivity due to nanostructuring at higher temperatures.

## I. INTRODUCTION

Phonon-glass electron-crystal [1] is a concept of ideal thermoelectric properties of crystal materials. Materials such as skutterudite [2], perovskite [3,4], and clathrate compounds [5,6] composed of cages and guest atoms are expected to realize this concept. In these materials, it is expected that phonons are significantly scattered by guest atoms while electrons can transport in the framework without significant scattering. Because a large free space of guest atoms is preferred to enhance phonon scattering by guest atoms [7,8], type-I clathrates [9] composed of Weaire-Phelan structure [10], which divides space into cage structures with a large volume, may be the most promising material. Indeed, they exhibit various intriguing phonon properties such as a glass-like plateau of temperature-dependent thermal conductivity, $\kappa(T)$, at low temperature [7,11,12], significant hardening of guest modes with temperature [13-15], and weak temperature

* ohnishi@photon.t.u-tokyo.ac.jp, shiomi@photon.t.u-tokyo.ac.jp

dependence of $\kappa(T)$ at high temperature [7,12,16].

Recent theoretical works on clathrates have revealed the basic effects of the guest atoms on phonon properties. The weak interaction between guest atoms and the framework results in flat bands, and the flat guest bands affect phonon properties in different manners. The main impact of the flat guest bands on phonon properties is to suppress phonon relaxation times in a wide range of frequencies [17-19]. It may be straightforward to understand that the flattening of bands increases scattering phase space (SPS) of phonons and decreases their relaxation time. In addition to SPS, the three-phonon coupling matrix represented by eigenvalues, polarization vectors, and atomic masses (described in Methodology in detail) among scattered phonons also has a vital role in decreasing the relaxation time according to detailed analyses on type-I $Ba_8Ga_{16}Ge_{30}$ (BGGe) [19]. As for phonon group velocity, while hybridization of the guest modes and framework modes decreases group velocities, theoretical works [18,19] have shown that the effect on group velocities is secondary. It has also been found that the glass-like plateau of temperature-dependent thermal conductivity at low temperature can be attributed to thermal excitation of phonons with higher frequencies than those of flat bands [20]. It may be interesting to note that the glass-like plateau is not a direct consequence of phonon anharmonicity because the effect of thermal excitation is relatively enhanced by the flat guest bands that suppress the contribution of low-frequency phonons to heat transport. Therefore, an in-depth understanding of the phenomena observed at high temperatures, e.g., the weak temperature dependence of thermal conductivity, should give us more knowledge on phonon anharmonicity.

In this study, we investigate the effects of strong phonon anharmonicity at high temperatures from first-principles calculations. Type-I clathrate $Ba_8Ga_{16}Sn_{30}$ (BGSn), which has a remarkably strong anharmonicity [21] compared with other off-center clathrates such as Sr- and Eu-filled Ge clathrates ($\{Sr/Eu\}_8Ga_{16}Ge_{30}$) [22], is mainly analyzed. Furthermore, type-I $Ba_8Ga_{16}Si_{30}$ (BGSi), which has much weaker anharmonicity than BGSn, is analyzed to highlight the effect of strong anharmonicity in BGSn. Quartic phonon anharmonicity is incorporated to obtain temperature-dependent *effective* second-order interatomic force constants (IFCs) using a self-consistent phonon (SCP) theory [23,24]. In addition to the phonon transport considered by the Peierls-Boltzmann transport equation (BTE), the coherent interbranch component is also considered by employing the so-called unified theory formulated by Simoncelli *et al*. [25]. These analyses reveal that lattice thermal conductivity can be increased by hardening of flat-band guest modes and the coherent thermal transport is not negligible above a few hundred degrees Kelvin. Because the hardening of guest modes significantly increases thermal conductivity at low frequencies, it leads to intriguing phenomena such as stronger suppression of thermal conductivity due to nanostructuring at higher temperatures.

## II. METHODOLOGY

### A. Interatomic Force Constants

Type-I clathrate BGSn with a highly symmetric structure ($Pm\overline{3}n$ space group) was analyzed in this study. For the sake of comparison, type-I BGSi, which has much weaker anharmonicity than type-I BGSn, was also analyzed. IFCs were obtained with first-principles calculations using the *Vienna ab initio simulation package* (VASP) [26]. Perdew-Burke-Ernzerhof exchange-correlation functional [27] with the projector augmented wave potential [28,29] was employed in all the calculations. The cutoff energies were set at 330 and 340 eV for BGSi and BGSn, respectively. A primitive unit cell containing 54 atoms was used as a model for all the simulations. The lattice constants of the clathrates were optimized using a $4 \times 4 \times 4$ Monkhorst–Pack $\boldsymbol{k}$-mesh, where $\boldsymbol{k}$ is the wavevector. The optimized lattice constants were 10.60 Å for BGSi and 11.90 Å for BGSn. While the former value is in excellent agreement with an experimental value (10.59 Å [30]), the latter is slightly larger than an experimental value (11.69 Å [31]). While the lattice thermal conductivity may depend on the value of the lattice constant [20], the values obtained through calculations of this study are used herein.

To obtain anharmonic IFCs, more than 120 randomly displaced structures were prepared, while harmonic IFCs were obtained through a finite displacement approach [32] with a displacement of 0.02 Å. First, molecular dynamics simulations were performed at a temperature of 500 K with a time step of 1.0 fs and a $2 \times 2 \times 2$ Monkhorst–Pack $\boldsymbol{k}$-mesh. The temperature was controlled using a velocity-scaling algorithm during the first few hundred femtoseconds of the simulation and then via the Nosé–Hoover algorithm [33]. After confirming that atomic vibrations had stabilized, the structures were extracted every 50 fs, and every atom was displaced along a random direction with a maximum magnitude of 0.1 Å to ensure that no two structures were correlated. Further, the anharmonic IFCs were calculated up to the sixth order using the least absolute shrinkage and selection operator solved using the coordinate descent method [34], while the harmonic IFCs were fixed with those obtained by the finite displacement method. Temperature-dependent harmonic IFCs were obtained by using the self-consistent phonon (SCP) theory [20,23] that modifies harmonic IFCs with the effect of quartic anharmonicity. The SCP calculations were performed at the Γ point ($\boldsymbol{q}$ = 0), and eigenvectors at arbitrary $\boldsymbol{q}$ points were obtained via Fourier interpolation. For all the simulations based on the SCP theory, off-diagonal components of self-energy and, thus, polarization mixing were considered in this study. In phonon dispersion of BGSn calculated with the obtained IFCs, the guest modes ($T_{2g}$ and $E_g$ modes), the lowest and second-lowest optical bands, are located around 20 $cm^{-1}$ at 300 K, as shown in Fig. S1. In addition to the guest modes, flat bands located around 40-50 $cm^{-1}$ are dominated by the framework, particularly Sn atoms.

### B. Lattice Thermal Conductivity

Using the obtained harmonic and cubic IFCs, lattice thermal conductivity without including the coherent part, $\kappa_{\mathrm{p}}$, was calculated by solving Peierls-BTE, which is associated with diagonal terms of group velocity operator [35]. Considering the second-order perturbation within the single-mode relaxation approximation, the linewidth due to the three-phonon scattering for phonon mode $q$ is derived as

$$\Gamma_q = \frac{\pi}{16N} \sum_{q_1,q_2} |V_3(-q,q_1,q_2)|^2 \left[(n_1+n_2+1)\delta\left(\omega_q-\omega_1-\omega_2\right) - 2(n_1-n_2)\delta\left(\omega_q-\omega_1+\omega_2\right)\right], \quad (1)$$

where the subscripts ($i$ = 1, 2) denote phonon modes contributing to the scattering of the target mode $q$, $n_i$ is the Bose-Einstein distribution function, $\omega_i$ is the phonon frequency, $N$ is the number of $\boldsymbol{q}$ points, and $\pm q = (\pm\boldsymbol{q}, s)$ with $\boldsymbol{q}$ and $s$ being the wavevector and branch index, respectively. The three-phonon coupling matrix element $V_3$ is given by

$$V_3(q,q_1,q_2)=\left(\frac{\hbar}{\omega\omega_1\omega_2}\right)^{\frac{1}{2}}\times\sum_{\boldsymbol{R}_i l_i p_i}\Psi^{p_0p_1p_2}_{0l_0,\boldsymbol{R}_1l_1,\boldsymbol{R}_2l_2}\times\frac{\boldsymbol{e}^{p_0}_{l_0}(q)\boldsymbol{e}^{p_1}_{l_1}(q_1)\boldsymbol{e}^{p_2}_{l_2}(q_2)}{\sqrt{M_{l_0}M_{l_1}M_{l_2}}}\times\exp[i(\boldsymbol{q}\cdot\boldsymbol{R}_0+\boldsymbol{q_1}\cdot\boldsymbol{R}_1+\boldsymbol{q_2}\cdot\boldsymbol{R}_2)], \quad (2)$$

where $\hbar$ is the reduced Planck constant, $\boldsymbol{R}_i$ is the position of the primitive cell, $l_i$ is the atom index, $p_i$ is the direction of the displacement of atom $l_i$, $M$ is the atomic mass, $\Psi$ is the cubic IFCs, and $\boldsymbol{e}(q)$ is the eigenvector of the mode $q$. The phonon lifetime due to three phonon scattering $\tau_{\mathrm{pp}}$ is given by $\tau_{\mathrm{pp}}(q)=1/(2\Gamma_q)$. The total phonon lifetime of mode $q$ is obtained with Matthiessen's rule: $\tau_q^{-1}=\tau_{q,\mathrm{pp}}^{-1}+\tau_{q,\mathrm{iso}}^{-1}+\tau_{q,\mathrm{b}}^{-1}$, where $\tau_{q,\mathrm{iso}}^{-1}$ is the scattering rate due to isotopes [36], and $\tau_{q,b}^{-1}=2|\boldsymbol{v}_q|/L_{\mathrm{g}}$ due to boundary of grains with an effective diameter $L_{\mathrm{g}}$, respectively. Finally, the Peierls term is calculated as $\kappa_{\mathrm{p}}^{\alpha\beta,\mathrm{scp}}(T)=(NV)^{-1}\sum_q c_q(T)v_q^{\alpha}(T)v_q^{\beta}(T)\tau_q(T)$ using IFCs modified with SCP theory. Here, $V$ is the volume of the primitive unit cell, $c$ is the mode specific heat, $v$ is the group velocity, and $\alpha$ and $\beta$ are the Cartesian directions. In this analysis, the result for $\alpha=\beta$ was taken because the type-I clathrates used were isotropic.

Because linewidths of phonons may be larger than interbranch spacings in some branches of BGSn, the contribution of interbranch coherent tunneling to heat transport may not be negligible. Therefore, the coherent component, corresponding to the nondiagonal terms of the heat flux operator, was calculated employing the unified theory [25,35]. Its contribution can be obtained as

$$\kappa_{\mathrm{c}}^{\alpha\beta}=\frac{\hbar}{k_B^2TVN}\sum_{\boldsymbol{q}}\sum_{s_1\neq s_2}\frac{\omega_1+\omega_2}{4}\mathrm{V}_{12}^{\alpha}(\boldsymbol{q})\mathrm{V}_{21}^{\beta}(\boldsymbol{q})\frac{\omega_1n_1(n_1+1)+\omega_2n_2(n_2+1)}{(\omega_1-\omega_2)^2+(\Gamma_1+\Gamma_2)^2}(\Gamma_1+\Gamma_2), \quad (3)$$

where $k_{\mathrm{B}}$ is the Boltzmann constant. The subscripts (1 and 2) denote the phonon modes $(\boldsymbol{q},s_1)$ and $(\boldsymbol{q},s_2)$, respectively. The generalized group velocity operator $\mathrm{V}(q)$ is given as

$$\mathrm{V}_{12}(\boldsymbol{q})=\frac{1}{\omega_1+\omega_2}\left\langle\boldsymbol{e}(q_1)\middle|\frac{\partial D(\boldsymbol{q})}{\partial\boldsymbol{q}}\middle|\boldsymbol{e}(q_2)\right\rangle, \quad (4)$$

where $D(\boldsymbol{q})$ is the dynamical matrix, which includes the anharmonic renormalization from the SCP theory.

### C. Electron Thermal Conductivity

Because electron contribution to thermal transport may not be negligible in BGSn above room temperature [37], electron thermal conductivity, $\kappa_{\mathrm{el}}$, was also calculated within the constant relaxation time approximation employing BoltzTrap2 [38]. Details are also documented in Sec. I of Supplementary Materials (SM). Electron band structures were calculated using VASP in basically the same manner as that for the phonon analysis. The $\boldsymbol{k}$-mesh density was set to $20\times20\times20$ for electron analysis, with which the convergence of electron properties such as electrical conductivity was confirmed. The carrier concentration is set to be an experimentally observed value, $3.2\times10^{18}$ cm$^{-3}$ [31], and the electron relaxation time was determined as 1.26 fs to reproduce the electrical conductivity at room temperature [31].

The principal reason to consider electron contribution is to reproduce an increase in thermal conductivity above room temperature due to the bipolar effect [37]. Although the temperature at which the bipolar effect becomes distinct strongly depends on the magnitude of the band gap, the accurate investigation of the band gap of clathrates is challenging. This difficulty comes from a unique feature of clathrates as well as a general problem of semilocal density functionals for exchange and correlation [39]. In the framework of clathrate compounds, atoms can be placed with different combinations of sites and elements, namely site occupancy factors, even with the same composition of elements, and their electronic states depend on the placement of atoms [40,41]. While it may be possible to estimate a plausible band gap by considering the effect of temperature and taking an ensemble average for possible structures [42-44] and carefully selecting an exchange-correlation functional, we simply modified the magnitude of band gap by manually adjusting the energy of the conduction bands in this analysis. This adjustment of the band gap modifies only the threshold temperature of bipolar effect, but not the increasing rate of electron thermal conductivity with temperature, as shown in Fig. S1 in SM.

## III. RESULTS AND DISCUSSIONS

Because type-I BGSn has a quadruple-well potential on Ba guest atom at 6*d* site, the quartic lattice anharmonicity and the associated temperature renormalization of phonon frequency are significant. Figure 1(a) shows a potential energy surface (PES) of the lowest optical mode at the Γ point ($T_{2g}$ mode), which is dominated by the guest atom. The deviation of the stable positions of the guest atom from the cage center was 0.69 Å that exceeded an experimental value of 0.43 Å [31]. This overestimation should be because of the larger lattice constant as mentioned in Sec. II A. The depth of the quadruple-well potential was 2.7 meV/atom, which corresponds to 31 K. Decomposed anharmonic contributions up to sixth-order show that even-order potentials dominate the potential on the guest atom at 6*d* site while the odd-order potentials become zero because of the symmetry. By employing the SCP theory, harmonic IFCs (blue dotted lines in Fig. 1(a)) were gradually renormalized by the quartic potential (red dash-dotted line) with increasing the temperature. While the SCP theory is effective in incorporating the finite-temperature effect, it is applicable only when atoms vibrate around a single position, center of a harmonic potential, but not around bottoms of a quadruple-well potential. For off-center systems, therefore, one can obtain harmonic potential being convex upward above a certain temperature, which was 53 K for the type-I BGSn. While eigenvalues have imaginary values without considering the effect of quartic anharmonicity, an inclusion of quartic anharmonicity lifts imaginary frequencies to positive values, as shown in Fig. 1(b). Following the hardening of phonon modes, the frequency of $T_{2g}$ mode is varied from 6.6 cm$^{-1}$ at 70 K to 16 cm$^{-1}$ at 300 K. The experimentally-measured frequencies of $T_{2g}$ mode are 15 cm$^{-1}$ at 70 K and 19 cm$^{-1}$ at 300 K [15]. This comparison shows that while the discrepancy between the experimental and theoretical value is distinct at low temperatures, it becomes smaller around room temperature. Calculation of partial density of states (DOS) shows that flat modes lying at low frequency can be divided into two groups, as shown in Fig. S2 in SM, which shows clearer phonon dispersion and DOS. The flat bands around 20 cm$^{-1}$ are dominated by Ba guest atoms while flat bands around 30-40 cm$^{-1}$ are dominated by the framework. We will show that phonon modes located at a frequency below the flat guest bands and phonon modes located between the flat guest and framework modes play an important role in realizing intriguing phonon properties of type-I BGSn.

Using the temperature-dependent harmonic IFCs and cubic IFCs, thermal conductivities of type-I BGSn were calculated. If the effect of quartic anharmonicity is considered, temperature-dependent thermal conductivity $\kappa(T)$ shows a peak even without considering any scattering process except for phonon-phonon scattering, as shown in Fig. 2. The peak temperature of $\kappa(T)$ was 59 K in this analysis. While there is no experimental data between 100 and 300 K, available experimental data shows that the peak temperature is at least above 100 K, which is higher than our estimation. The underestimation of the peak temperature should be because of the underestimation of frequencies of flat guest bands that decreases the temperature at which the effect of thermal excitation of higher-frequency phonons than flat bands becomes distinct.

At higher temperatures (> 300 K), calculated values are in good agreement with experimental data. Our result shows (a) the absence of the temperature dependence of lattice thermal conductivity excluding the coherent contribution ($\kappa_{\mathrm{p}}^{\mathrm{scp}}$: blue line in Fig. 2) between 300 and 600 K, (b) significant contribution of coherent component ($\kappa_{\mathrm{c}}$: orange lines), and (c) a sharp increase in thermal conductivity because of electronic contribution ($\kappa_{\mathrm{el}}$: green lines). To reproduce the increase of thermal conductivity above 300 K observed in the experiment [37], the electronic band gap was modified from 0.045 to 0.15 eV. While the increase of $\kappa_{\mathrm{el}}$ might be noticeable, its effect simply can be attributed to the bipolar effect (see Fig. S1 in SM). We will, therefore, focus on more intriguing phenomena of phonon transport below.

To gain insight into the slow decay of $\kappa(T)$ with temperature and absence of the temperature dependence above the peak temperature (> 100 K), spectral and cumulative $\kappa_{\mathrm{p}}^{\mathrm{scp}}(T)$ of type-I BGSi and BGSn are compared in Fig. 3. In usual single-crystal materials, including type-I BGSi, that do not have strong phonon anharmonicity, modal thermal conductivity decreases with increasing temperature at most frequencies because of the enhancement of phonon-phonon scattering, as shown in Fig. 3(a). In type-I BGSn, surprisingly, thermal conductivities at low frequency (below 50 cm$^{-1}$) increases with the temperature above the peak temperature, as shown in Fig. 3(b). Accumulated $\kappa_{\mathrm{p}}^{\mathrm{scp}}(T)$ up to 50 cm$^{-1}$ was 0.11, 0.14, and 0.19 Wm$^{-1}$K$^{-1}$ at 100, 300, and 600 K that correspond to 25%, 49%, and 70% of the total $\kappa_{\mathrm{p}}^{\mathrm{scp}}(T)$, respectively. This increase of thermal conductivity of low-frequency phonons compensates the decrease of heat transport of higher-frequency phonons (> 50 cm$^{-1}$), leading to the absence of the temperature dependence. Thermal conductivity divided into different frequency ranges show an almost opposite feature in BGSi and BGSn, as shown in Fig. S3. In BGSi, low-frequency phonons dominate the total heat transport at low temperatures, and its contribution relatively decreases slightly compared with the high-frequency contribution. In BGSn, however, while high-frequency phonons dominate the total heat transport at around the peak temperature (60-100 K), the contribution of low-frequency phonons increases as increasing temperature. For example, the contribution of phonons below 40 cm$^{-1}$ to the total $\kappa_{\mathrm{p}}^{\mathrm{scp}}$ decreases from 68% at 100 K to 58% at 300 K for BGSi while it increases 22% at 100 K to 45% at 300 K for BGSn.

The increase in the contribution of low-frequency phonons with temperature may lead to an intriguing phenomenon from an engineering point of view. Polycrystallization is a technique to reduce phonon mean-free-path (MFP) and, thus, thermal conductivity for thermoelectric materials [45-47]. In materials that do not have strong anharmonicity, the reduction of thermal conductivity due to polycrystallization is larger at a lower temperature because of longer phonon MFPs. For example, the reduction of $\kappa_{\mathrm{p}}^{\mathrm{scp}}$ due to the introduction of 5-μm effective grains is larger at lower temperature, as shown in Fig. 4(a). However, for BGSn, the reduction of thermal conductivity is weak at around the peak temperature of $\kappa(T)$ because of the large contribution of high-frequency phonons. The reduction of $\kappa_{\mathrm{p}}^{\mathrm{scp}}$ due to 5-μm grains is 13% at 100 K, 18% at 300 K, and 19% at 600 K. This finding gives designing strategies for phonon engineering. While it may be difficult to tune phonon MFP in materials with strong anharmonicity at around the peak temperature, the tunability may improve at higher temperatures.

To gain a deeper insight into the increase in thermal conductivity at low frequency with temperature, phonon properties were analyzed in more detail. Figure 3(b) shows that low-frequency phonons have different temperature dependence in 0-20 cm$^{-1}$ and 30-40 cm$^{-1}$. Wider-frequency phonons contribute to heat transport at higher temperatures in the former range, and thermal conductivity increases suddenly between 300 and 600 K in the latter range. The dip frequency of the spectral $\kappa_{\mathrm{p}}^{\mathrm{scp}}(T)$ of type-I BGSn around 10-20 cm$^{-1}$ clearly corresponds to the frequency range of flat guest modes, which is highlighted in Fig. S4. Figures 5(a) and 5(b) show that the flat guest modes, which are explicitely at 8-20 cm$^{-1}$ at 100 K, 16-24 cm$^{-1}$ at 300 K, and 22-28 cm$^{-1}$ at 600 K (see Fig. S4), have small lifetimes and group velocities. (One can see an aggregations of data shifting to a higher frequency at the corresponding frequencies with increasing temperature.) Figure 5(a) also shows that phonon lifetimes at 30-40 cm$^{-1}$ suddenly increase from 300 to 600 K.

To investigate effects of the change in lifetimes and group velocities due to the hardening of guest modes, their effects were analyzed by calculating a virtual thermal conductivity: $\kappa_{\mathrm{p},q}^{\mathrm{scp}}(T_c, T_v, T_\tau) = c_q(T_c)\boldsymbol{v}_q^2(T_v)\tau_q(T_\tau)$. Each property, namely $c_q$, $\boldsymbol{v}_q^2$, and $\tau_q$, was calculated with the renormalized phonon dispersion, including the polarization vector and frequency, by the SCP theory and the Bose-Einstein distribution at a given temperature. The effect of a specific phonon property can be analyzed by changing the corresponding temperature in $\kappa_{\mathrm{p},q}^{\mathrm{scp}}(T_c, T_v, T_\tau)$. The corresponding phonon modes at different temperatures were identified to minimize the sum of inner products of all the corresponding eigenvectors at each $\boldsymbol{q}$-point. Figures 5(c) and 5(d) show changes in the modal thermal conducitvity, $\kappa_{\mathrm{p},q}^{\mathrm{scp}}(T_c, T_v, T_\tau)$, due to each phonon property at intermidiate temperatures (from 100 to 300 K) and high temperatures (from 300 to 600 K), respectively. As shown in Fig. 5(c), the increase in thermal conductivity at 0-20 cm$^{-1}$ at the intermediate temperature range can be attributed to an increase in group velocities. Indeed, with increasing temperature from 100 to 300 K, $\kappa_{\mathrm{p}}^{\mathrm{scp}}$ at 0-20 cm$^{-1}$ is increased by 44% because of the increase in group velocities. While the total $\kappa_{\mathrm{p}}^{\mathrm{scp}}$ decreases because of the decrease in lifetimes at high frequency (> 60 cm$^{-1}$), the increase in group velocities makes the thermal conductivity decay slower with respect to temperature than the Klemens model, $\kappa_{\mathrm{p}}(T) \propto T^{-1}$ [48]. While group velocities still increase above 300 K at 10-20 cm$^{-1}$ because of the hardening of guest modes, the effect of the increase in lifetimes at 30-40 cm$^{-1}$ (between flat bands dominated by the guest atoms (≈ 20 cm$^{-1}$) and framework (40-50 cm$^{-1}$) as shown in Fig. S1) overcomes the increase in group velocity, as shown in Fig. 5(d). Consequently, this abnormal increase in phonon lifetimes at a higher frequency than that of the guest modes leads to the absence of temperature dependence of $\kappa(T)$.

To gain more insights into the abnormal increase in lifetimes at high temperatures, we focused on a phonon mode whose lifetime increased the most. First, we identified that the lifetime of a phonon mode at $\boldsymbol{q}$ = (0.22, 0.22, 0.22) and $\omega$ = 36 cm$^{-1}$ at 300 K (see Fig. S4) increased the most from 300 to 600 K (from 1.2 to 17.5 ps). Figures 5(e) and 5(f) show, respectively, the magnitude of the three-phonon coupling matrix element, $|V_3|^2$ term, which is determined by eigenvectors, frequencies, and atomic masses of the scattered phonons (see Eq. 2) and the $|V_3|^2$ term multiplied by SPS term at 300 and 600 K. These figures clearly show that the decrease in $|V_3|^2$ leads to the decrease in the scattering rate and, thus, the increase in the lifetime of the target phonon mode.

Finally, we have analyzed the contribution of coherent thermal transport. Considering this effect, we have successfully reproduced the magnitude of thermal conductivity of type-I BGSn above room temperature, as shown in Fig. 2. Figure 6 shows that flat bands at 20-60 $cm^{-1}$ mainly contribute to the coherent thermal transport at around the peak temperature (100 K). In particular, flat bands of the framework around 40-50 $cm^{-1}$ have a large impact. The figure also shows that coupling between nearly-degenerate states ($\omega_1 \simeq \omega_2$) has a large contribution in this low-frequency range (20-60 $cm^{-1}$). However, because phonons at this low-frequency range are excited up to 100 K, $\kappa_{\mathrm{c}}(T)$ at 20-60 $cm^{-1}$ does not change significantly with increasing temperature above 100 K. Instead of the low-frequency phonons, the contribution of phonon pairs with different frequencies ($\omega_1 \neq \omega_2$) at higher frequency (> 60 $cm^{-1}$) increases with temperature. For example, the contribution of phonons above 60 $cm^{-1}$ increases from 28% at 100 K to 49% at 600 K. In the total lattice thermal conductivity ($\kappa_{\mathrm{p}}^{\mathrm{scp}} + \kappa_{\mathrm{c}}$), the contribution of the coherent heat transport increases from 12% at 100 K to 24% at 600 K, which is not negligible.

## IV. CONCLUSION

In conclusion, we have revealed the complicated effects of strong quartic anharmonicity on phonon properties of a type-I clathrate with quadruple-well potential. The whole lattice contribution, including interbranch coherent contribution, was analyzed for lattice contribution by employing first-principles anharmonic lattice dynamics combined with the SCP theory and the unified theory. Above around the peak temperature of temperature-dependent thermal conductivity, $\kappa(T)$, $\kappa(T)$ has been successfully reproduced by considering electron contribution as well as the lattice contributions. It was confirmed that the sharp increase in thermal conductivity observed in an experiment can be attributed to electron contribution. It was also found that the contribution of the coherent thermal transport is not negligible at high temperatures (24% at 600 K). More interestingly, we found that lattice thermal conductivity of phonons located at a lower frequency than those of flat bands of the framework (40-50 $cm^{-1}$) increases with temperature, and this increase leads to the weak temperature dependence or a slight increase in the total lattice thermal conductivity with increasing temperature.

Detailed analysis showed that the low-frequency region (< 40 $cm^{-1}$) can be further divided into two regions by the flat guest modes ($T_{2g}$ and $E_g$ mode around 20 $cm^{-1}$). Above the peak temperature of $\kappa(T)$, the contribution of low-frequency phonons located below the frequency of the guest modes increases with temperature because of the increase in group velocities following the hardening of the guest modes. While this effect continues above room temperature, the contribution of higher-frequency phonons located above those of the guest modes increases more significantly because of the abnormal increase in their lifetimes. This abnormal increase in the contribution of low-frequency phonons with temperature can be attributed to the decrease in the three-phonon matrix elements, $|V_3|$ terms, with guest modes. From an engineering point of view, the increase in thermal conductivity of low-frequency phonons may lead to a larger decrease in thermal conductivity due to nanostructuring at higher temperatures. These findings indicate that it may be possible to manipulate phonon properties by tunning phonon anharmonicity of materials by using substitutions or vacancies, which can pave a new path of phonon engineering.

## V. ACKNOWLEDGEMENTS

The numerical calculations in this work were carried out on the facilities of the Supercomputer Center, Institute for Solid State Physics, the University of Tokyo and MASAMUNE-IMR at Center for Computational Materials Science, Institute for Materials Research, Tohoku University (Project No. 20S0514).

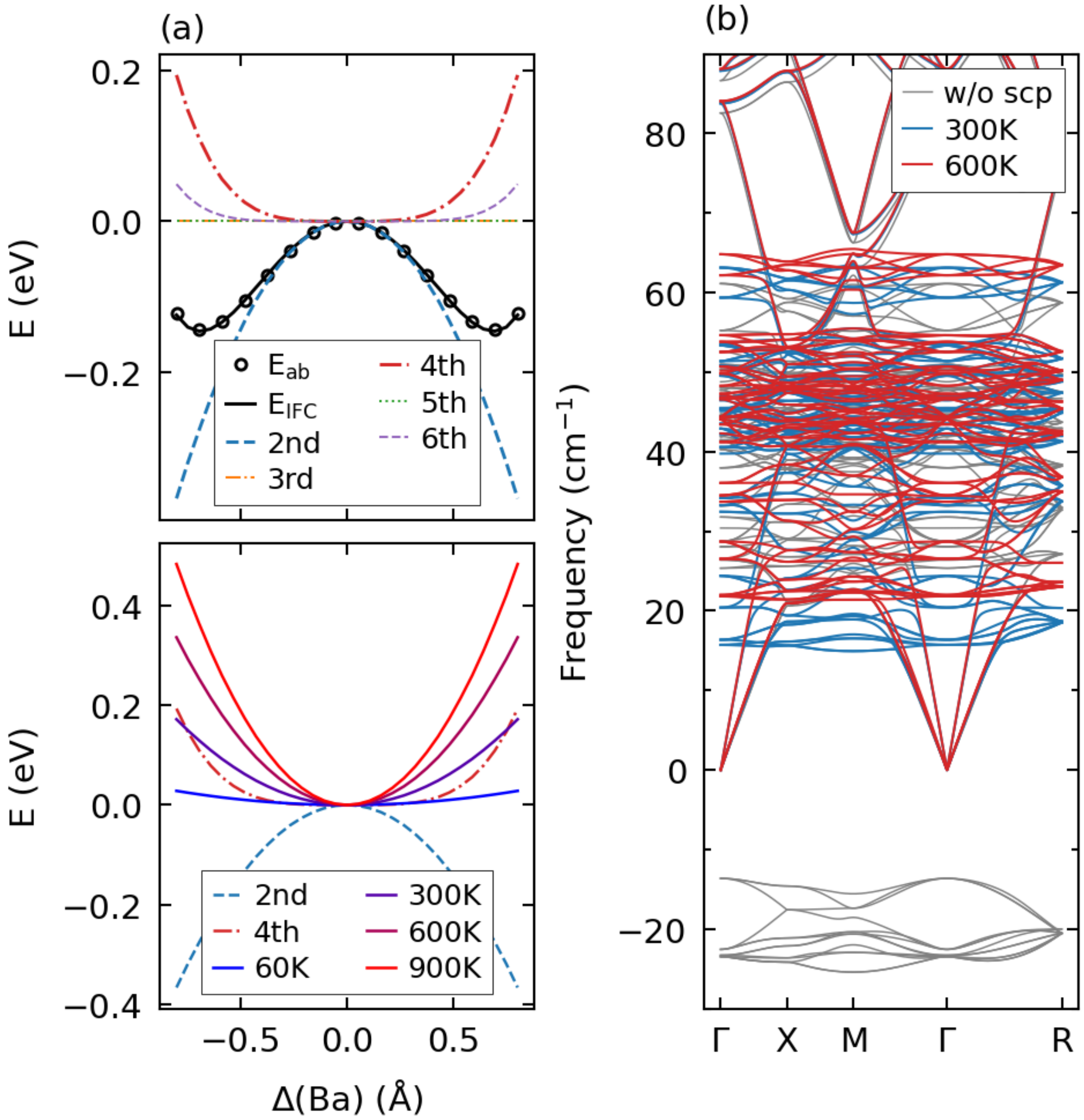

Fig. 1. Temperature-dependent harmonic phonon properties of type-I BGSn. (a) Potential energy surface (PES) of the lowest guest mode at the Γ point ($T_{2g}$ mode). In the top panel, black circles show values obtained directly from first-principles calculations and the black line shows data obtained through the interatomic force constants. Colored lines show PESs decomposed up to sixth-order. In the bottom panel, solid lines show temperature-dependent harmonic potentials. (b) Phonon dispersions obtained without (grey) and with (blue and red) considering the quartic anharmonicity. The latter is shown for 300 and 600 K.

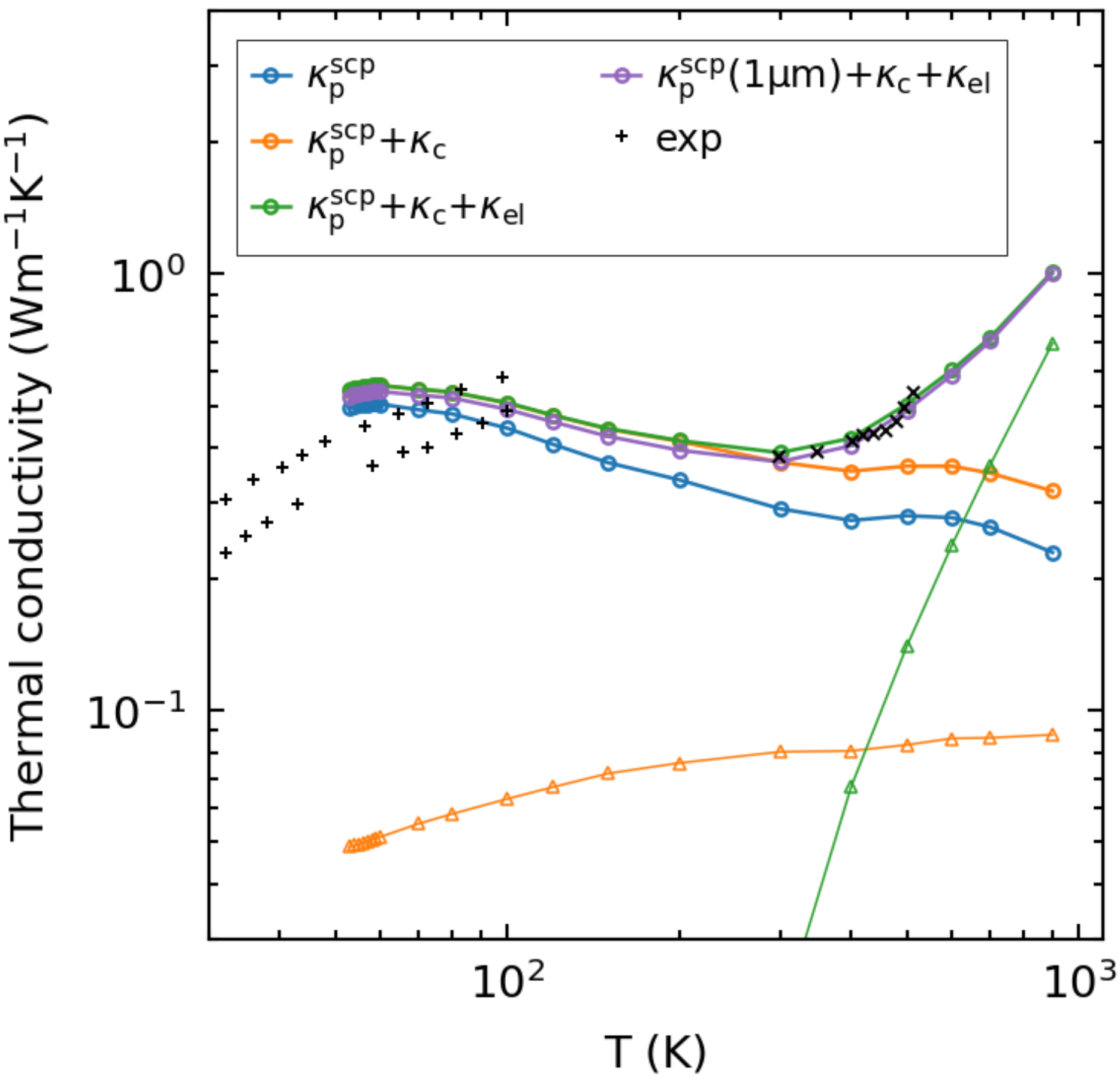

Fig. 2. Temperature-dependent thermal conductivity of type-I BGSn. Blue, orange, green, and purple lines denote the nondiagonal Peierls contribution ($\kappa_p^{scp}$), interbranch coherent contribution ($\kappa_c$), electron contribution ($\kappa_{el}$), and the effect of grain boundaries with the effective size of 1 μm, respectively. Circles show integrated values for different contributions while triangles show values for each contribution. Crosses show experimental data of single crystals obtained from Ref. [31] for temperatures below 100 K and Ref. [37] above 300 K.

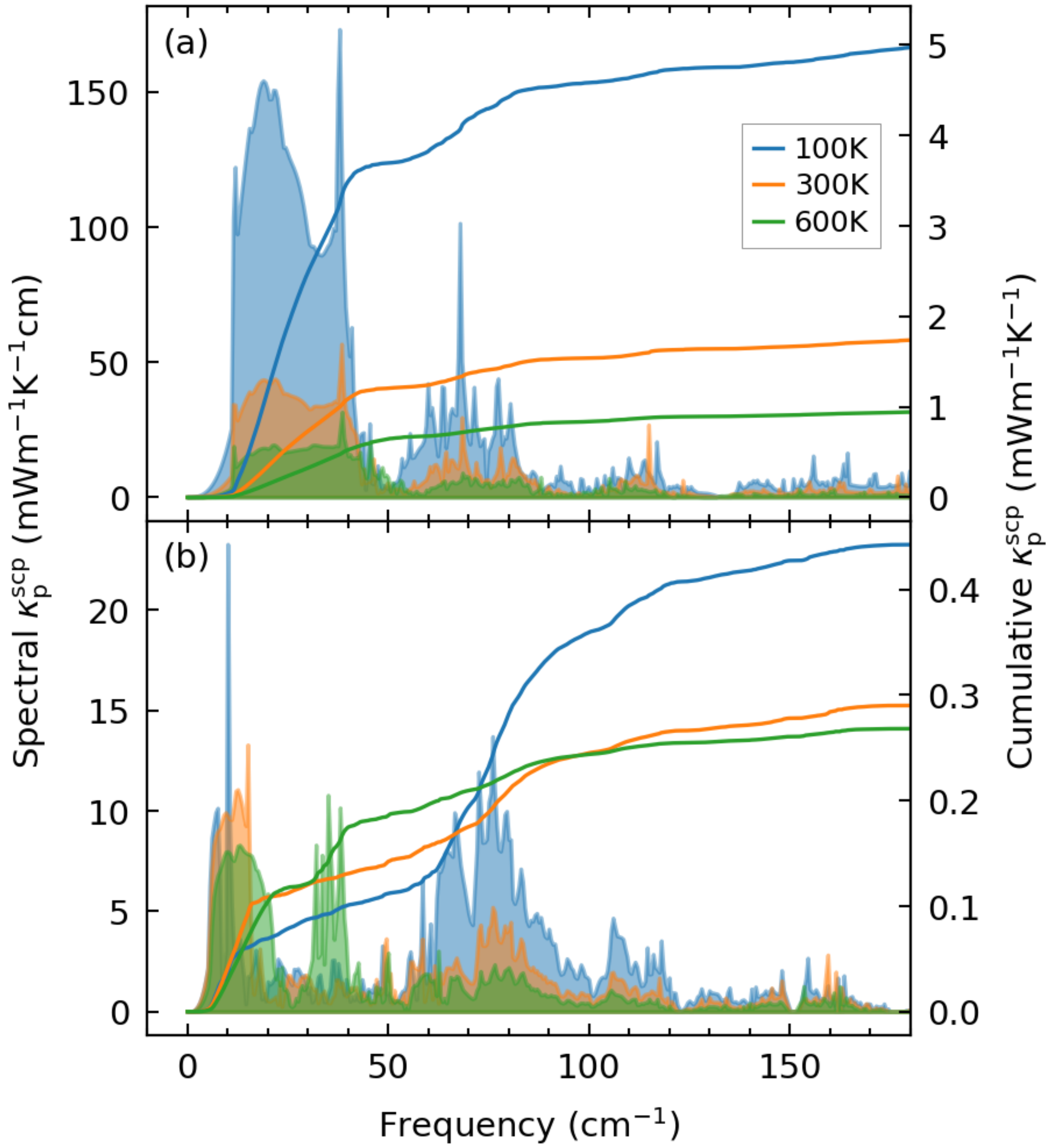

Fig. 3. Spectral and cumulative lattice thermal conductivities of (a) BGSi and (b) BGSn at different temperatures. The contribution of interbranch coherent component is not considered.

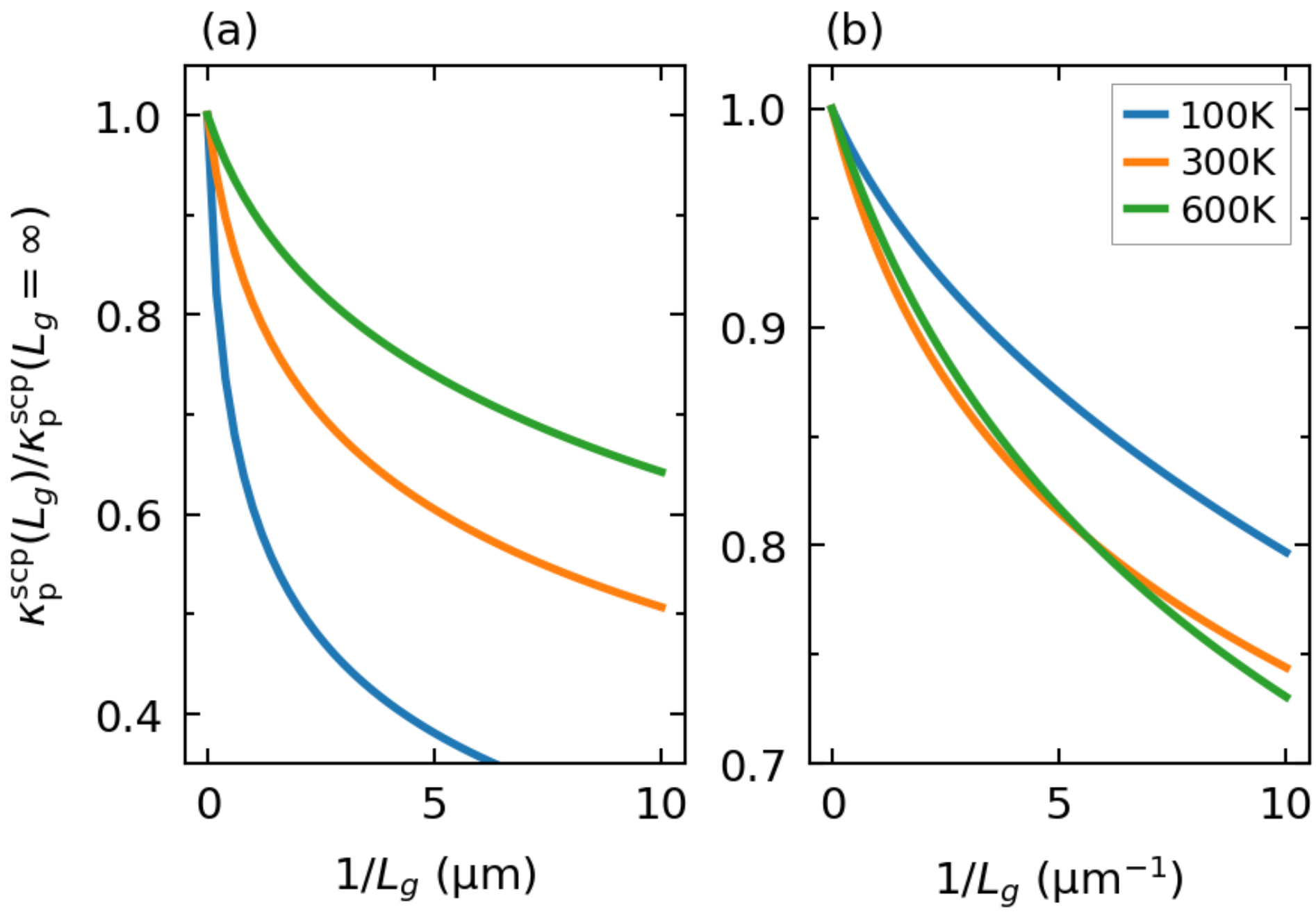

Fig. 4. Decrease in thermal conductivity due to polyscyrtallization for type-I (a) BGSi and (b) BGSn.

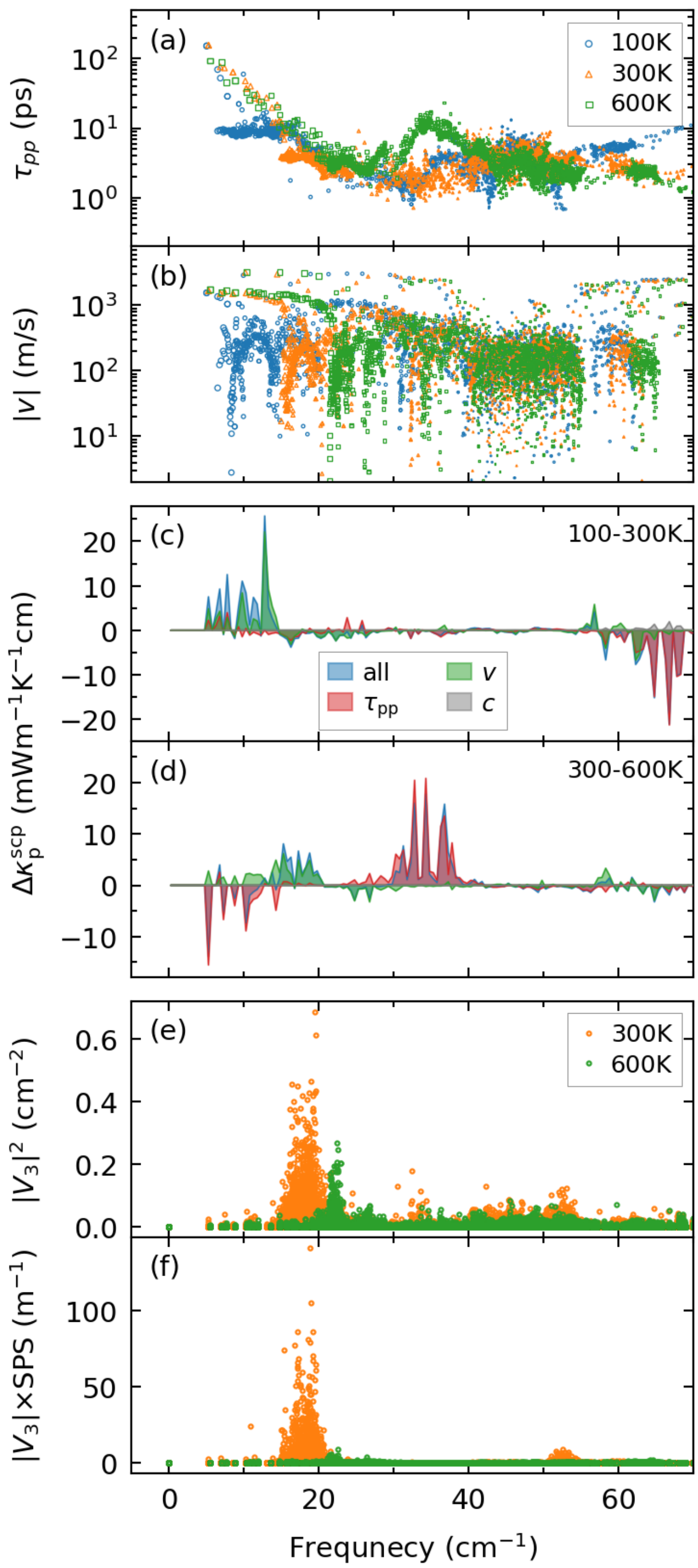

Fig. 5. Effect of hardening of flat guest bands in type-I BGSn. (a) Lifetime due to phonon-phonon scattering and (b) group velocity at different temperatures. (c), (d) Change in modal thermal conductivities due to different phonon properties (group velocity, lifetime, and other factors) from (c) 100 to 300 K and (d) 300 to 600 K. Note that phonon frequencies depend on temperature in the analysis with the SCP theory. In panel (c) and (d), phonon frequencies at 100 and 300 K are taken, respectively. (e), (f) The magnitude of the three-phonon coupling matrix element, $|V_3|^2$, (e) and $|V_3|^2$ multiplied by the SPS term (f) of the mode whose lifetime increases the most between 300 and 600 K (from 1.2 to 17.5 ps) that locates at 36 cm$^{-1}$ and $\boldsymbol{q}$ = (0.22, 0.22, 0.22) at 300 K (see Fig. S4).

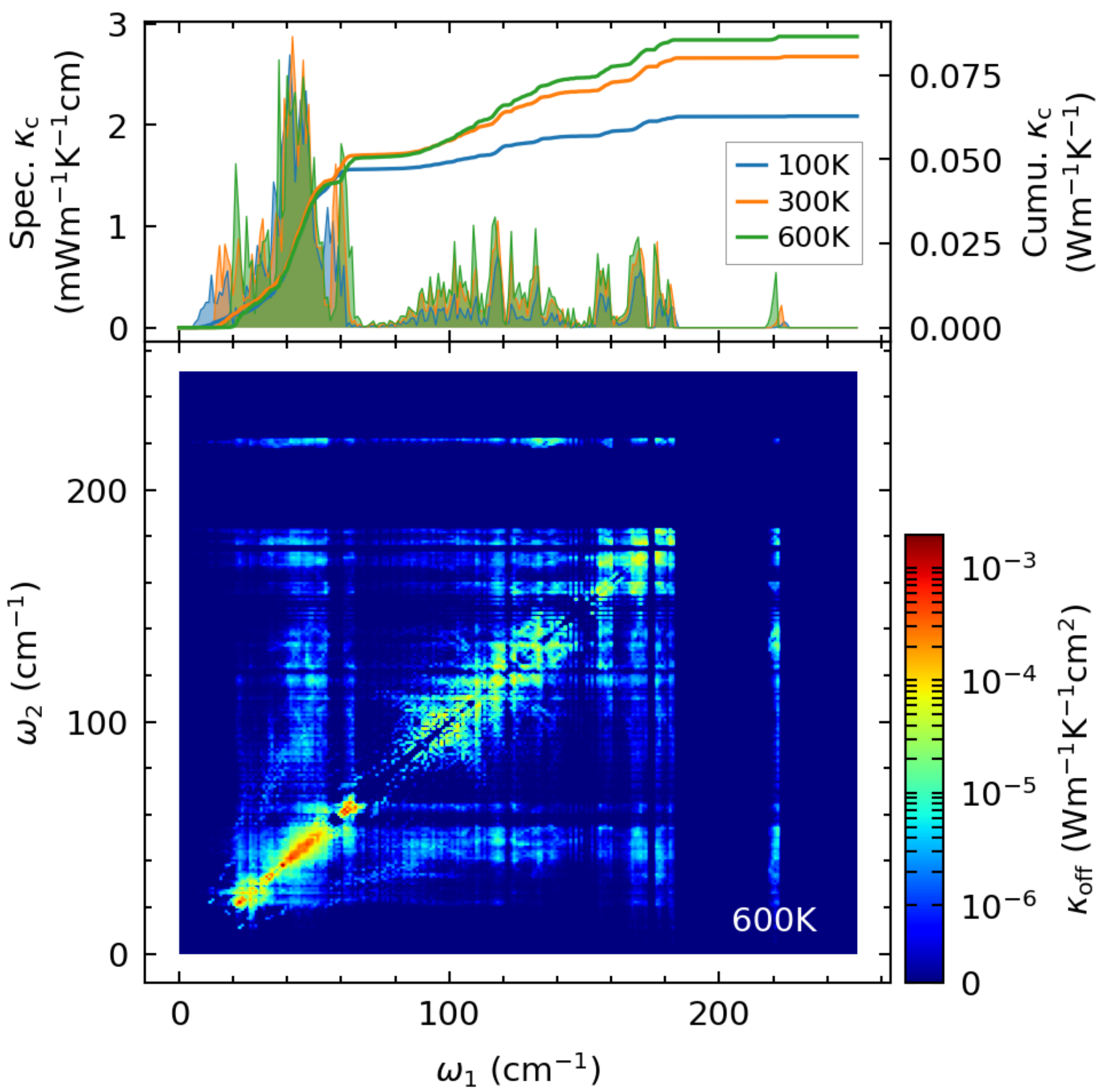

Fig. 6. Contribution of the interbranch coherent component, $\kappa_{\mathrm{c}}$, of type-I BGSn. The top panel shows spectral and cumulative thermal conductivities at different temperatures while the bottom panel shows the two-dimensional modal thermal conductivity, $\kappa_{\mathrm{c}}(\omega_1, \omega_2)$ (see Eq. (3)). To obtain the spectral and cumulative thermal conductivities (top), the contribution of a mode $q_1$ of a coupling between two modes $q_1$ and $q_2$ is distributed as $c_{q_1}/(c_{q_1} + c_{q_2})$, where $c_q$ is the mode specific heat.